\documentclass[lettersize,journal]{IEEEtran}
\usepackage{amsmath,amsfonts}
\usepackage{algorithmic}
\usepackage{algorithm}
\usepackage{array}
\usepackage[caption=false,font=normalsize,labelfont=sf,textfont=sf]{subfig}
\usepackage{textcomp}
\usepackage{stfloats}
\usepackage{url}
\usepackage{verbatim}
\usepackage{graphicx}
\usepackage{cite}
\usepackage{booktabs} 
\usepackage{multirow} 

\begin{document}

\title{Transformer-Based Model for Cold Start Mitigation in FaaS Architecture}

\author{Alexandre Savi Fayam Mbala Mouen,
Jerry Lacmou Zeutouo,
Vianney Kengne Tchendji~\IEEEmembership{Member, IEEE}
\thanks{Manuscript received XX, 20XX; revised XX, 20XX.}
\thanks{Alexandre S.F.M.M. and Vianney K.T. are with Departement of Mathematics and Computer Science, University of Dschang, Dschang, Cameroon}
\thanks{Jerry L.Z. is with the MIS Labotory, University of Picardie Jules Verne, 33 rue Saint-Leu, 80039 Amiens Cedex, France.}

}

\markboth{IEEE Transactions on XXX}%
{Fayam \MakeLowercase{\textit{et al.}}: Transformer-Based Model for Cold Start Mitigation in FaaS Architecture}


\maketitle

\begin{abstract}
Serverless architectures, particularly the Function as a Service (FaaS) model, have become a cornerstone of modern cloud computing due to their ability to simplify resource management and enhance application deployment agility. However, a significant challenge remains: the cold start problem. This phenomenon occurs when an idle FaaS function is invoked, requiring a full initialization process, which increases latency and degrades user experience. Existing solutions for cold start mitigation are limited in terms of invocation pattern generalization and implementation complexity. In this study, we propose an innovative approach leveraging Transformer models to mitigate the impact of cold starts in FaaS architectures. Our solution excels in accurately modeling function initialization delays and optimizing serverless system performance. Experimental evaluation using a public dataset provided by Azure demonstrates a significant reduction in cold start times, reaching up to 79\% compared to conventional methods. 
\end{abstract}

\begin{IEEEkeywords}
serverless, FaaS, cold start, deep learning, time-series forecasting, Transformers, latency, performance optimization
\end{IEEEkeywords}

\section{Introduction}
\IEEEPARstart{T}{he} rapid emergence of cloud computing has transformed the digital industry by enabling immediate and flexible access to virtualized computing resources. Services such as storage, computing, and networking allow organizations to reduce costs \cite{noble_machine_2023} while benefiting from scalable, on-demand infrastructure managed by external providers. At the core of this evolution is the Function as a Service (FaaS) model, an innovative solution that allows functions to be deployed in response to specific events without requiring the management of underlying resources. This model, foundational to serverless architectures, provides increased flexibility and usage-based billing, reducing both costs and operational complexity for developers. Serverless architectures, built upon the FaaS model, abstract infrastructure management, allowing developers to focus solely on the code and features of their applications \cite{gias_cocoa_2020}. This approach optimizes application scalability and simplifies infrastructure management, which is a notable advantage in an era of massive automation. Organizations increasingly seek streamlined processes to enhance responsiveness and efficiency, making serverless architectures a key driver of modern digital transformation.

However, despite its numerous advantages, FaaS poses specific challenges \cite{shahrad_serverless_2020}, particularly the cold start issue. Cold start latency arises when functions must be loaded into containers from an idle state \cite{vahidinia_mitigating_2023}. This occurs when a function has not been invoked for some time, necessitating a complete reinitialization of the runtime environment—a step that can introduce significant delays. While tolerable in certain applications, cold starts become problematic in high-performance or latency-sensitive environments. Cold start impacts user experience and the responsiveness of serverless applications, especially in scenarios with highly variable workloads. This limitation represents a major barrier to the widespread adoption of FaaS for latency-sensitive applications, necessitating robust solutions to mitigate its effects and improve serverless platform performance.

In recent years, various studies have focused on optimizing resource management and reducing cold starts in FaaS environments. However, most existing work relies on static allocation techniques or heuristic workload analyses, which fail to account for the highly dynamic nature of FaaS platforms. Moreover, few solutions leverage advancements in artificial intelligence (AI) and deep learning, which offer powerful tools to predict workloads and proactively manage resource needs. To address these limitations, our research proposes a deep learning-based approach, specifically using Transformers, to effectively mitigate cold start delays and frequency in FaaS architectures. The choice of Transformers is justified by their ability to capture complex relationships within temporal data, enabling precise function usage predictions and proactive resource management. By integrating a Transformer-based model, our approach aims to reduce function startup delays while optimizing resource consumption.

This paper makes several significant contributions, including :
\begin{enumerate}
    \item Implementation of a Transformer-based model for cold start mitigation, incorporating usage predictions that facilitate proactive resource allocation. Results demonstrate substantial improvements in latency reduction and prediction accuracy, validated by standard evaluation metrics ;
    \item Performance evaluation through simulations on the OpenWhisk FaaS platform and tests on real-world datasets. Our experiments show improvements in prediction accuracy, cold start delay reduction, and resource optimization, confirming the validity and effectiveness of our approach.
\end{enumerate}

The remainder of this paper is organized as follows :
\begin{itemize}
    \item Section II – FaaS \& Cold Start : Provides an in-depth exploration of the Function as a Service concept, its advantages for application deployment, and the cold start limitations inherent in this architecture.
    \item Section III – Literature Review : Analyzes existing research on workload prediction and cold start mitigation, discussing common resource management approaches and identifying the limitations of traditional methods.
    \item Section IV – Proposed Approach : Introduces our predictive approach to cold start mitigation, explaining the use of deep learning models, particularly Transformers, to address this challenge.
    \item Section V – Evaluation : Describes our evaluation methodology and experimental results, analyzing the performance of our model on real-world and simulated datasets.
    \item Conclusion : Summarizes the contributions of our research, discusses the impact of our results, and proposes future research directions.
\end{itemize}

\section{FaaS \& Cold Start}
With the rapid evolution of cloud computing, the serverless paradigm has revolutionized the way modern applications are designed and deployed. It enables businesses and developers to delegate the management of the underlying infrastructure to cloud service providers while focusing solely on developing the business logic of their applications. Among the various serverless models, Function as a Service (FaaS) has emerged as one of the most promising due to its ability to execute functions in response to events \cite{eismann_state_2022}, reduce operational costs, and provide automatic scalability. FaaS is an innovative approach that allows developers to deploy and run computing units as standalone functions in a cloud environment. These functions are typically triggered by external events \cite{nguyen_managing_2023,eismann_state_2022}, such as HTTP requests, database changes, or IoT (Internet of Things) events. Unlike traditional cloud computing models such as Infrastructure as a Service (IaaS) or Platform as a Service (PaaS), where developers must provision and manage servers or complete environments, FaaS completely abstracts this management, enabling instant deployment and execution of functions on demand.

The core principle of FaaS is based on three main characteristics :
\begin{itemize}
    \item Automatic Elasticity : FaaS is designed to dynamically adapt to workload changes. When demand for a function increases, the cloud provider automatically provisions additional resources to handle the new requests, ensuring applications scale seamlessly with traffic fluctuations ; Conversely, when demand decreases, resources are automatically released.
    \item Usage-Based Billing : unlike traditional models where billing is based on server or virtual machine allocation time, FaaS billing is directly tied to the actual runtime of functions \cite{noauthor_cncf_2018}. This approach significantly reduces costs, as developers pay only for the CPU time and memory used during function execution. For example, in scenarios where functions are invoked sporadically, costs are minimized ;
    \item No Infrastructure Management : the FaaS model simplifies development by eliminating the need for developers to manage servers, operating systems, or network configurations. These tasks are handled entirely by the cloud service provider, enabling developers to focus solely on function logic. This reduces deployment time and streamlines DevOps processes by removing much of the operational complexity associated with infrastructure management.
\end{itemize}
	
These features have driven the widespread adoption of FaaS in environments where scalability, flexibility, and cost-effectiveness are priorities, such as in microservices-based applications. In the FaaS model, each function is triggered by a specific event. A function is a stateless computing unit that executes precise logic and returns a result in response to that event. Unlike other cloud paradigms, FaaS does not store data between function executions. This stateless behavior is essential for automatic scaling, as it allows each request to be processed in isolation, independently of others.

Figure 1 illustrates the general process of executing a function in a FaaS architecture \cite{golec_cold_2023,jegannathan_time_2022}:

\begin{figure}
    \centering
    \includegraphics[width=1\linewidth]{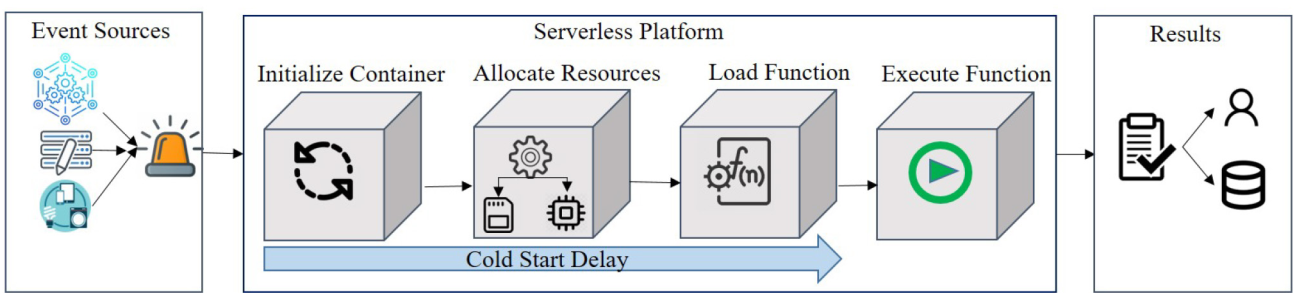}
    \caption{FaaS execution flow \cite{vahidinia_mitigating_2023}}
    \label{fig:enter-label}
\end{figure}

\begin{enumerate}
    \item Event Triggering : an event from an external source, such as an HTTP request or an IoT notification, is captured by the cloud provider ;
    \item Container Initialization : if no active instance of the function exists, the system provisions a new container to execute the function ;
    \item Dependency Loading : the function code and associated libraries are loaded into the container. This process may take time depending on the function's complexity and the volume of required dependencies ;
    \item Function Execution : once the container is ready, the function executes and returns the result to the event source (e.g., an end user via an HTTP API) ;
    \item Resource Release : at the end of the function's execution, the container and all its resources will be released. This is known as scale-to-zero \cite{mampage_deep_2023, kumari_acpm_2023}, an innovative feature in FaaS that helps reduce costs \cite{tirkey_novel_2023}.
\end{enumerate}

This sequence of tasks is transparent to the developer but has a direct impact on performance, particularly when addressing cold starts. Cold start is one of the primary challenges in FaaS architectures. It occurs when a function is invoked for the first time or after a long period of inactivity. In such scenarios, the cloud service provider must allocate a new container to host the function and initialize the runtime environment, resulting in additional latency before the function can respond to requests \cite{barcelona-pons_benchmarking_2021, matta_function-as--service_2021, steinbach_tppfaas_2022}. Cold starts can significantly impact the overall performance of applications in FaaS architectures, especially for latency-sensitive applications or those with real-time high-availability requirements \cite{tirkey_novel_2023, manner_cold_2018}. Some scenarios where cold start can pose a problem include :
\begin{itemize}
    \item Real-Time Applications : in applications such as financial transaction processing systems or emergency alert systems, even a few milliseconds of latency can have significant consequences. In these contexts, cold starts lasting several seconds can compromise user experience and overall system performance ;
    \item Demand Spikes : when an application experiences a sudden traffic surge, cold starts can occur if existing containers cannot handle the new demand. This results in increased latency as new containers must be provisioned to handle the growing demand ;
    \item Public APIs : for widely used public APIs, excessive latency caused by a cold start can lead to longer response times, potentially affecting thousands of simultaneous users .
\end{itemize}

Cold start is essentially a zero-scaling problem, where a function instance is decommissioned to save resources when the function is not in use. While this ensures efficient resource usage, restarting the function from scratch requires a certain delay before it becomes ready again. The latency of a cold start varies based on factors \cite{lee_mitigating_2021} such as container size, dependency volumes, and the specific configuration of the cloud platform. For example, a study conducted by measured cold start delays for several popular cloud providers, including AWS Lambda, Google Cloud Functions, and Azure Functions. The results showed that cold starts could take anywhere from 100 milliseconds to several seconds, depending on the complexity of the functions and the service provider. FaaS is undoubtedly one of the most transformative innovations in cloud computing, offering unprecedented flexibility and scalability. However, like any emerging technology, it comes with challenges, particularly the issue of cold starts. This challenge can significantly affect application performance, particularly in environments where latency and responsiveness are critical. Proposed solutions, such as pre-allocated instances, code optimization, and the adoption of lighter runtimes, offer ways to mitigate these impacts. However, continued research is needed to improve these mechanisms and ensure optimal performance in large-scale FaaS architectures.

\section{Literature Review}
FaaS (Function-as-a-Service) architectures, despite their advantages in scalability and simplified resource management, face the challenge of cold starts. This phenomenon, characterized by the delay required to initialize an idle function, can significantly impact the performance of serverless applications, especially when low latency is critical. This section explores approaches developed to mitigate cold starts, focusing on strategies to reduce the delay and frequency of this phenomenon. We review techniques ranging from container pre-warming to machine learning approaches, highlighting their quantitative outcomes, practical constraints, and limitations.

\subsection{Techniques for Reducing Cold Start Delay}
Cold start delay reduction techniques aim to decrease the time required for functions to be ready to handle requests. This delay can vary depending on container configuration, code volume, and specific dependencies. To address this, several approaches have been proposed in the literature:
\begin{itemize}
    \item Container Pre-Warming: Platforms such as OpenWhisk \cite{noauthor_apacheopenwhisk_2024} implement pre-warmed container strategies. The primary idea is to maintain a pool of “ready” containers in memory. Upon a function invocation, a pre-warmed container is immediately assigned, significantly reducing startup delays. OpenWhisk, for instance, uses two pre-warmed containers with a Node.js runtime, achieving startup delay reductions of up to 30\% in certain cases, especially for simple functions without heavy dependencies. However, this approach has practical drawbacks. It requires maintaining a pool of containers continuously, leading to resource wastage during low or unpredictable demand. Additionally, it is limited to environments configured for specific programming languages, reducing flexibility.
    \item Periodic Function Invocation: To keep functions active and reduce startup delays, platforms like CloudWatch \cite{noauthor_aws_nodate-1} and Lambda Warmer \cite{noauthor_lambda-warmer_2023} employ periodic function invocation. These tools trigger functions at regular intervals, irrespective of actual demand, to maintain “warm” containers. This technique reduces startup delays by 20\% to 40\% for intermittently used applications. However, the primary downside is that it does not adapt to workload variations and may lead to significant resource wastage during prolonged inactivity.
    \item Selective Code Optimization - FaaSLight \cite{liu_faaslight_2023}: introduced by Liu et al., proposes loading only the essential parts of application code to reduce startup latency. By initializing only the necessary code, this technique achieves a startup latency reduction of up to 25\%. However, this approach is more suitable for applications with simple architectures and minimal dependencies. For more complex applications requiring multiple modules and dependencies, its effectiveness may be limited.
    \item Function Fusion \cite{lee_mitigating_2021}: A novel approach by Lee et al. involves fusing sequential functions into one. This technique avoids a cold start for the second function by combining both into a single container, reducing the overall workflow delay. Experiments have shown an 18\% reduction in startup latency for sequential applications. However, for scenarios requiring parallel functions, this approach may result in extended execution times, limiting its applicability.
    \item Proactive Prediction with SARIMA \cite{jegannathan_time_2022}: Advanced methods like time-series prediction models, such as SARIMA, coupled with an autoscaler, predict workloads based on invocation history to initialize containers ahead of peak demand. This approach reduces startup delays by 22\% to 30\% in high-demand environments. However, it relies heavily on accurate prediction data and may be ineffective in the face of unexpected workload changes.
\end{itemize}
In summary, these techniques offer notable reductions in function startup delays but have limitations in terms of implementation complexity and operational costs. For instance, container pre-warming and periodic invocation require additional resources and are challenging to adapt to highly variable demand environments.

\subsection{Techniques for Reducing Cold Start Frequency}
Techniques aimed at reducing cold start frequency focus on preserving containers in memory to minimize the number of cold starts during repeated invocations. This strategy is vital for serverless applications requiring rapid and frequent responses.
\begin{itemize}
    \item Temporary Retention of Active Containers: Platforms such as AWS Lambda \cite{noauthor_aws_nodate-1} and Google Cloud Functions  \cite{noauthor_cloud_nodate} keep containers in a paused state for several minutes after function execution. If a new request arrives during this period, the already-in-memory container is reused, eliminating startup delays and reducing cold start frequency by 15\% to 20\%. However, this technique increases memory usage and requires optimal thresholds for container release. Poor configurations may result in additional costs without significant benefits.
    \item Reinforcement Learning with Q-Learning  \cite{agarwal_reinforcement_2021, agarwal_deep_2023}: Using reinforcement learning algorithms like Q-learning, it is possible to dynamically predict the optimal number of function instances based on demand. By adjusting container retention frequency based on usage patterns, this technique significantly reduces cold starts while minimizing CPU resource usage. Studies have shown a reduction in cold starts by approximately 40\%, with up to 55\% CPU savings. However, Q-learning requires extensive training data and can be complex to adapt for generalized applications.
    \item Hybrid Histogram Policy : The Hybrid Histogram Policy introduced by Shahrad et al.  \cite{shahrad_serverless_2020}is an adaptive approach that dynamically adjusts the duration containers remain active. Using histograms to determine function invocation frequencies and optimal inactivity periods, this policy reduces cold start frequency by up to 35\%, especially for functions with irregular demand. However, it is challenging to implement and requires real-time invocation data for precise model adjustments.
    \item Workload Prediction with LSTM Neural Networks: Kumari et al. \cite{kumari_mitigating_2022} proposed using LSTM neural networks to predict the number of containers to keep active and the optimal retention duration based on workload patterns. Results indicate a reduction in cold start frequency by 18\% to 25\%, significantly improving performance for serverless applications with variable demand. However, this approach requires a robust infrastructure to process real-time predictions and extensive historical data for training, limiting its accessibility for certain organizations or projects.
    \item In another study similar to Kumari's, Vahidinia et al. \cite{vahidinia_mitigating_2023} use a reinforcement learning method and LSTM models to determine the number of warm containers and the duration for which they are kept warm.
\end{itemize}
While these approaches effectively reduce cold start frequency, they are often resource-intensive and costly. Temporary container retention and machine learning-based methods require continuous monitoring and adjustments to remain effective. Although machine learning and reinforcement learning methods offer promising solutions, resource optimization is necessary to ensure practical adoption in large-scale environments.

\section{Proposed Approach}
In Function-as-a-Service (FaaS) architectures, cold start is a significant performance obstacle, primarily because it increases latency during the first invocation of a function after a period of inactivity. To mitigate this issue, our study proposes an approach based on Transformers tailored for time-series data, following the architecture developed by Kashif \cite{rasul_kashifpytorch-transformer-ts_2024} and its specific improvements for temporal data. Transformers provide a distinct advantage over traditional recurrent models (such as LSTM and GRU) by managing temporal dependencies without requiring sequential computations, thereby accelerating training and improving prediction accuracy.

By leveraging an attention mechanism particularly effective for managing long-term dependencies, our model adapts to the variability of workloads in FaaS environments, ensuring optimal resource management and reducing cold starts. The primary innovations introduced by the Transformer model adapted for time-series data include its ability to capture complex relationships and generate probabilistic forecasts, offering a more comprehensive and tailored view of future demands. The choice of Transformers, specifically Kashif’s model tailored for time-series, is justified for several reasons :
\begin{itemize}
    \item Handling Long-Term Dependencies : unlike LSTMs, Transformers can efficiently capture long-range relationships through their attention mechanism. This is critical in time-series data, where long-term dependencies significantly influence workload prediction ;
    \item Parallelized Computation : transformers exploit parallelized operations, making training faster and more suitable for large datasets—especially advantageous for FaaS platforms, which analyze vast invocation histories ;
    \item Probabilistic Forecasting : by incorporating probabilistic predictions, our model estimates the uncertainty associated with each forecast. This is essential in cloud environments where precise and well-calibrated forecasts are necessary to avoid resource overprovisioning.
\end{itemize}
	
The Transformer model for time-series data by Kashif is designed to process sequentially structured data while integrating multiple feature types (statistical, temporal, contextual) for each data point. The methodology is broken down as follows :
\begin{enumerate}
    \item Encoding Time-Series Data : each data point is enriched with a set of temporal and static features :
    \begin{itemize}
        \item pastValues : Historical values of the time series that provide the historical context of the function.
        \item pastTimeFeatures : temporal characteristics for each \texttt{pastValues} point, such as day of the month or month of the year, enabling the model to detect seasonal patterns or recurrences.
        \item staticCategoricalFeatures : static information associated with the series (e.g., function or client ID), adding fixed context to help the model distinguish between multiple functions or applications.
    \end{itemize}

    \begin{figure}
        \centering
        \includegraphics[width=1\linewidth]{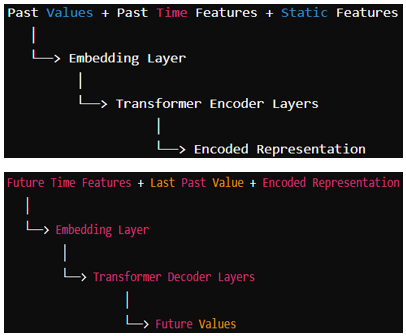}
        \caption{Input Representation of the Encoder and Decoder}
        \label{fig:enter-label}
    \end{figure}

    \item Attention Mechanism : the core of the Transformer model lies in the attention mechanism. Rather than processing data sequentially, attention assigns weights to points in the sequence based on their relevance to the prediction. This mechanism is vital for detecting nonlinear patterns and capturing interactions across different time periods:
    \begin{itemize}[leftmargin=1.5cm]
        \item Multi-Head Attention : each attention layer includes several "heads" that can focus on different parts of the time series simultaneously, capturing various aspects of temporal patterns.
        \item Positional Encoding : since Transformers lack inherent sequential structure like RNNs, positional encoding is added to maintain the sequence order. This encoding helps the model distinguish each point's position within the sequence.
    \end{itemize}

    \item Model Architecture : the model consists of two main blocks: the encoder and the decoder, specifically modified for time-series forecasting. The architecture draws inspiration from the vanilla Transformer originally designed for natural language processing (NLP). However, instead of processing sequences of words through semantic embeddings, our model adapts the inputs to handle numerical time-series data, as illustrated in the accompanying figure 2. In NLP, inputs are token embeddings combined with positional encodings. Here, pastValues, pastTimeFeatures, and staticCategoricalFeatures replace textual embeddings. These features are projected into a latent space via dedicated embedding layers, enabling the model to capture temporal dynamics (e.g., historical trends, seasonality) and static context (e.g., function identifiers). The encoder fuses these representations to generate a global historical context, while the decoder leverages futureTimeFeatures and the last observed value to predict future invocations. This adaptation of inputs, combined with multi-head attention, allows the model to process time-series data with the same flexibility as the vanilla Transformer processes sentences, but tailored to numerical and cyclical patterns inherent to FaaS workloads.
    \begin{itemize}[leftmargin=1.5cm]
        \item Encoder :
        \begin{itemize}[leftmargin=1cm]
            \item Input : the encoder takes \texttt{pastValues}, \texttt{pastTimeFeatures}, and \texttt{staticCategoricalFeatures}.
            \item Processing : these inputs are combined to produce an encoded representation summarizing the historical context. Multi-head attention captures long-range temporal dependencies, offering a holistic view of the sequence.
        \end{itemize}
        \item Decoder :
        \begin{itemize}[leftmargin=1cm]
            \item Input : the decoder uses \texttt{futureTimeFeatures} (temporal characteristics of each future point) and the last value of \texttt{pastValues}.
            \item Processing : using the encoded representation and \texttt{futureTimeFeatures}, the decoder generates \texttt{futureValues} (predicted values). These predictions incorporate identified temporal patterns and contextual information provided by static and temporal features.
        \end{itemize}
    \end{itemize}

    \item Learning Latent Representations : unlike LSTMs, which struggle to generalize across multiple series, Kashif’s Transformer model learns latent representations for a wide range of time-series data. It captures data distributions, enabling probabilistic and non-point predictions (i.e., forecasts represented as distributions).

    \item Probabilistic Forecasting : the model does not produce a single forecast but captures a distribution of potential outcomes, estimating uncertainty in the predictions \cite{noauthor_blogtime-series-transformersmd_nodate}. This feature is crucial for adjusting container management based on probabilities of future demand.
\end{enumerate}

\section{Evaluation}
This section details the experimental results of our Transformer model for proactive cold start management in FaaS environments. Two essential dimensions were examined: reducing cold start latency and minimizing the frequency of cold starts. The evaluation is accompanied by a direct comparison with LSTM models, highlighting the significant improvements that Transformers offer in FaaS use cases.

\subsection{Experimental Framework and Model Configuration}
A central focus of our experimentation was the handling of HTTP-triggered functions. We based our work on a dataset from Microsoft Azure Functions, collected in July 2019\cite{noauthor_azurepublicdatasetazurefunctionsdataset2019md_nodate,shahrad_serverless_2020}. This dataset originally contained a wide variety of function invocations, but we specifically filtered out only those triggered by HTTP calls. This choice was motivated by two key factors:
\begin{itemize}
    \item HTTP-triggered functions are the most unpredictable, displaying large fluctuations in invocation rates ;
    \item They are also among the most popular invocation patterns in serverless platforms.
\end{itemize}
To address the two critical cold start dimensions—latency (i.e., the “delay” problem) and frequency (i.e., how often cold starts occur)—we leveraged clustering to better characterize the different invocation behaviors. By focusing on popular/frequently used functions for the cold start “delay” problem and irregular/low-usage functions for the cold start “frequency” problem, we ensured that our model could handle a wide spectrum of workload patterns in FaaS.

\subsubsection{Data preprocessing and clustering with DBSCAN (Figure 3)}
In our initial data processing phase, we started with 14 separate log files, each representing one day’s worth of Azure Functions invocations. We first filtered the data to keep only HTTP-triggered functions (identified via the "Trigger" field). These 14 files were then merged into a single dataset for uniform processing. To capture the variety of invocation patterns, we applied the DBSCAN (Density-Based Spatial Clustering of Applications with Noise) algorithm. This step was crucial for segmenting the data into clusters of functions that exhibited similar invocation patterns. Through this process, we identified 33 distinct clusters of HTTP-triggered functions. Each cluster corresponds to a unique pattern of invocations over time, ranging from highly frequent to very sparse usage. From each cluster, we then selected representative functions for subsequent analysis. 

\begin{figure}
    \centering
    \includegraphics[width=1\linewidth]{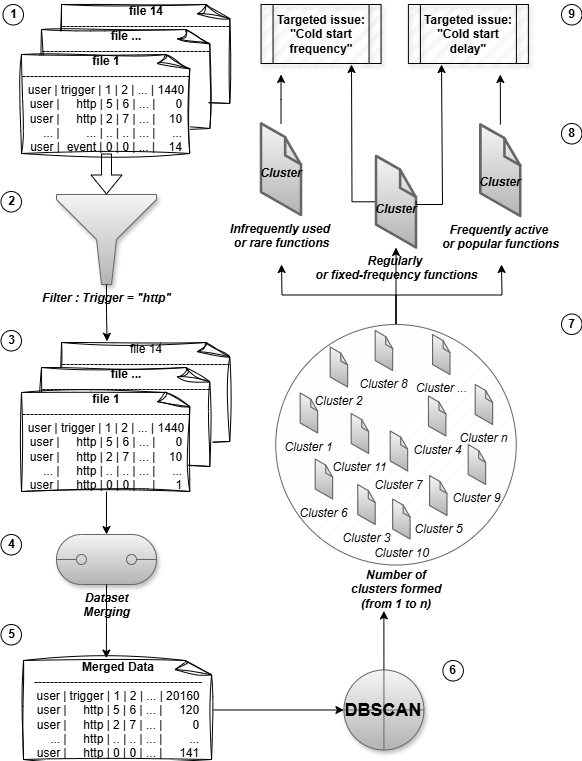}
    \caption{Global Data Processing Pipeline for Training the Transformer Model}
    \label{fig:enter-label}
\end{figure}

This clustering strategy directly supports the cold start objectives:
\begin{itemize}
    \item High- or Frequent-Demand Functions : often subject to cold start delay concerns, because predicting their peak loads can help us proactively prepare the runtime environment.
    \item Irregular or Low-Demand Functions : more relevant to cold start frequency issues, since these functions often remain idle for long periods, leading to more frequent container spin-ups.
    \item Regular or Fixed-Frequency Functions : interestingly, some functions showed stable and unchanging invocation intervals. Their patterns were so predictable that applying either cold start “delay” or “frequency” mitigation strategies consistently yields very high accuracy.
\end{itemize}

\subsubsection{Transformer model setup and hyperparameters}
After segmenting the data into meaningful clusters, we trained a TimeSeriesTransformers model to forecast invocation patterns. The goal was twofold:
\begin{enumerate}
    \item Predict the timing and volume of incoming requests for popular or frequently used functions (mitigating cold start delay by proactively preparing containers).
    \item Predict the intervals between function calls for less frequent or highly irregular functions (mitigating cold start frequency by intelligently adjusting container idle windows).
\end{enumerate}
Key hyperparameters and configurations for our Transformer model included:
\begin{itemize}
    \item Contextual Window (contextLength) : 200 points, giving the model a substantial historical view to understand trends and seasonalities.
    \item Prediction Length (predictionLength) : 100 points, ensuring the model forecasts sufficiently far into the future to anticipate upcoming demand spikes or lulls.
    \item Number of Encoder/Decoder Layers 
 : 4 layers each, enabling the model to capture both short-term fluctuations and longer-term dependencies.
    \item Dimension of the Model (dModel) : 32, specifying the size of the hidden representations used in multi-head attention.
    \item Embedding Dimension (embeddingDimension) : 2, helping to encode categorical features or cluster identifiers.
    \item Lags Sequence / Cardinality : additional parameters tailored to the time-series nature of the data, enabling the model to incorporate various seasonalities or repeating patterns.
\end{itemize}

We conducted the model training on Google Colab, leveraging its GPU acceleration for faster computation:
\begin{itemize}
    \item Python 3 (GPU) runtime, utilizing hardware acceleration for large-scale training.
    \item System RAM: ~12.7 GB, sufficient for loading and processing merged invocation logs.
    \item GPU RAM: ~15.0 GB, enabling the training of deeper Transformer networks without running out of memory.
\end{itemize}
For comparison, we trained a standard LSTM model with a similar number of layers and comparable hidden dimensions. This ensures that the improvement we observe with Transformers is primarily due to the architectural benefits—such as multi-head attention—rather than differences in training data size or compute resources.

\subsubsection{Proactive Cold Start management with OpenWhisk (Figure 4)}
To evaluate and deploy our model, we integrated the Transformer-based predictions into an Apache OpenWhisk serverless platform. OpenWhisk natively provides prewarmed containers that are ready to handle incoming requests, mitigating some cold start penalties. However, the default configurations can be somewhat static or purely reactive, which may not suffice for sudden demand spikes or highly irregular usage patterns. Our approach modifies and enhances OpenWhisk’s internal settings:
\begin{itemize}
    \item Proactive Allocation of Prewarmed Containers : by predicting high-demand intervals for frequently invoked functions, we increase the pool of prewarmed containers in advance. This directly addresses the cold start delay problem by ensuring containers are ready when requests arrive.
    \item Adaptive Container Inactivity Window : for functions that are invoked infrequently, we leverage the model’s forecasts of inter-arrival times. This allows us to adjust the “container idle time” or “window of container inactivity” dynamically, reducing the cold start frequency for sporadic invocations.
\end{itemize}

\begin{figure*}[t]
    \centering
    \includegraphics[width=0.8\textwidth]{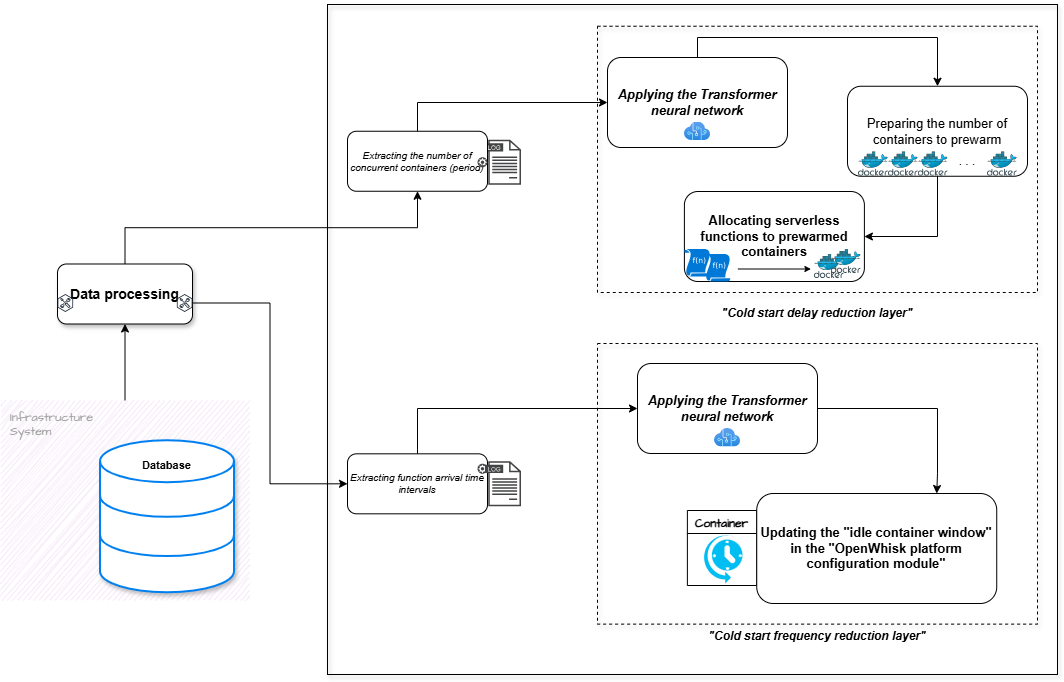}
    \caption{Proposed Approach: Transformer-Based Model to Mitigate Cold Start}
    \label{fig:enter-label}
\end{figure*}

By combining the Transformers forecasting capabilities with OpenWhisk’s container management, we create an adaptive FaaS platform that addresses both cold start delay and frequency in a unified framework. This configuration ensures minimal resource waste—only provisioning additional containers when truly needed—and near-instant response for popular functions, all while mitigating frequent cold starts for low-usage functions.

\subsection{Reducing Cold Start Delay}
The cold start delay represents the initialization time required when a function is triggered from an idle state. By optimizing startup times, the Transformer model enhances responsiveness and reduces function latency, ensuring a better user experience.

To achieve this, we rely on historical invocation data to predict future demand. In our broader dataset of HTTP-triggered functions, we randomly selected 12 different invocation patterns—each referred to as a dataset—to evaluate how well our model can predict and mitigate cold starts. Two main data scenarios were tested:
\begin{itemize}
    \item Minute-by-minute Invocations (Figure 5 : Original Data) : this scenario uses the original, unaltered logs where functions are invoked continuously, potentially every minute. Such fine-grained data more closely resembles real-world serverless workloads with unpredictable bursts and lulls.
    \begin{figure*}
        \centering
        \includegraphics[width=1\linewidth]{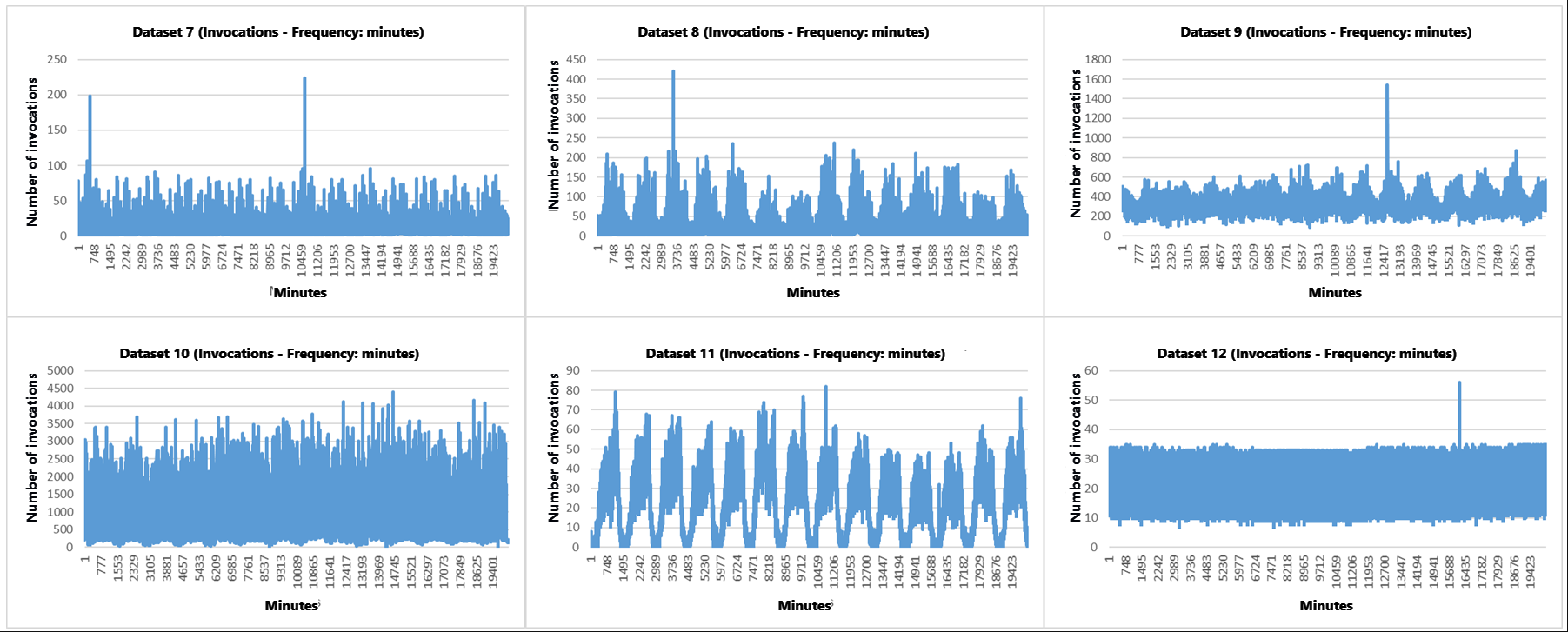}
        \caption{Original Data : Minute-by-minute Invocations}
        \label{fig:enter-label}
    \end{figure*}
    \item Hourly-aggregated Invocations (Figure 6 : Altered Data) : here, invocations are grouped by hour, simulating a coarser granularity similar to certain datasets found in the literature. While less granular, this scenario allows us to compare our model’s performance against references in other research works and to see how it handles simpler patterns.
    \begin{figure*}
        \centering
        \includegraphics[width=1\linewidth]{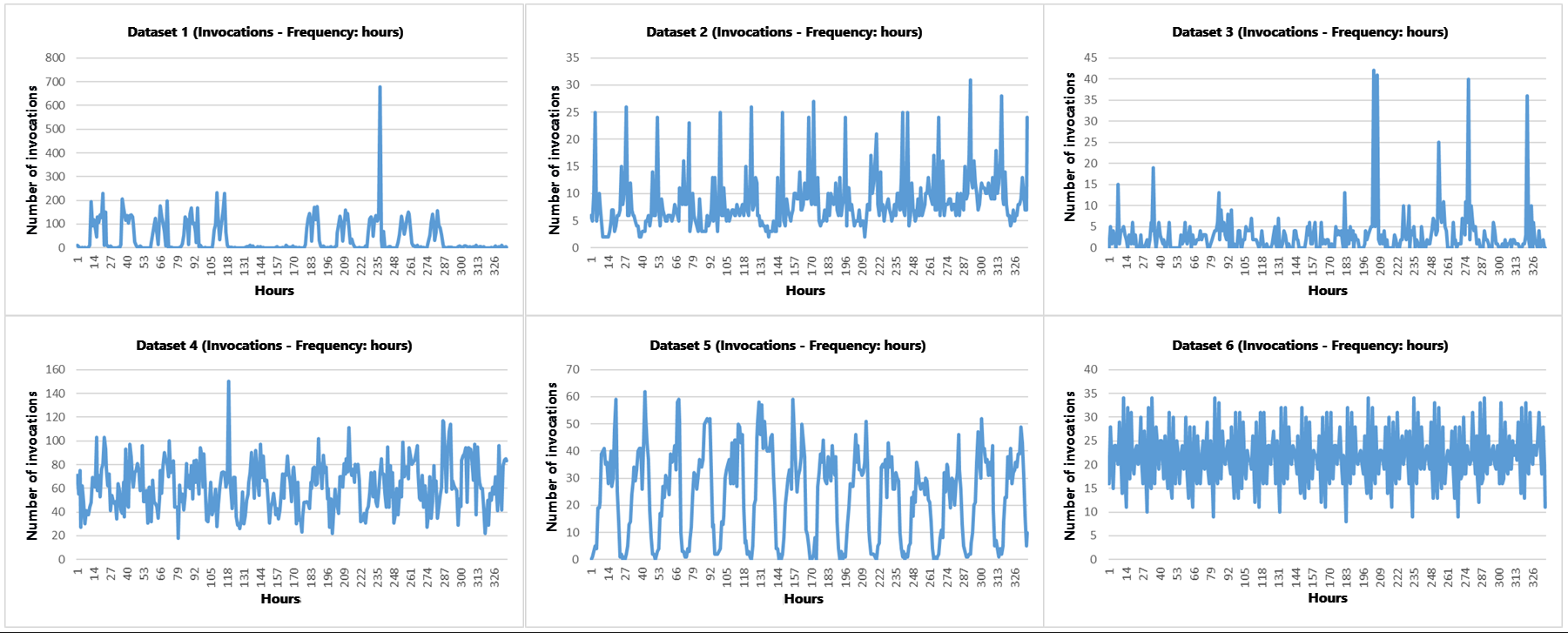}
        \caption{Altered Data : Hourly-aggregated Invocations}
        \label{fig:enter-label}
    \end{figure*}
\end{itemize}

We compared the Transformer to a standard LSTM model across multiple standard metrics—such as sMAPE (Symmetric Mean Absolute Percentage Error), Explained Variance, RMSE (Root Mean Squared Error), Normalized RMSE, R² Score, and Spearman Correlation, to capture different facets of prediction error and variance :
\begin{itemize}
    \item Tables I \& Table II illustrate performance results for cold start delay predictions, examining both hourly and minute invocation frequencies.
    \item Minute-Level Predictions: The Transformer achieved a ower sMAPE compared to LSTM (e.g. : Dataset 12 : Transformer sMAPE = 0.043 vs. LSTM = 0.108).
    \item Hourly Predictions: for altered datasets, the Transformer demonstrated superior stability, with explained variance up to 0.958 (e.g. : Dataset 12), indicating robust capture of workload variability.
    \item Normalized RMSE: The Transformer reduced errors in high-variability scenarios (e.g. : Dataset 10: Transformer N. RMSE = 0.101 vs. LSTM = 0.119).
\end{itemize}

\begin{table}[h]
\centering
\caption{Performance Results - Cold Start Delay: Dataset (Frequency: Hour)}
\begin{tiny} 
\begin{tabular}{llcccccc}
\toprule
\textbf{Functions} & \textbf{Model} & \textbf{sMAPE} & \textbf{E.V.\textsuperscript{a}} & \textbf{RMSE} & \textbf{N. RMSE\textsuperscript{b}} & \textbf{R2 Score} & \textbf{S. C.\textsuperscript{c}} \\
\midrule
\multirow{2}{*}{Dataset 1} & LSTM & \textbf{1.204} & -3.195 & 8.468 & 0.705 & -5.607 & -0.034 \\
& Transformers & 1.611 & \textbf{-0.781} & \textbf{3.659} & \textbf{0.365} & \textbf{-0.876} & \textbf{0.065} \\
\midrule
\multirow{2}{*}{Dataset 2} & LSTM & \textbf{0.321} & 0.202 & 4.641 & 0.193 & 0.196 & \textbf{0.499} \\
& Transformers & 0.358 & \textbf{0.710} & \textbf{3.874} & \textbf{0.161} & \textbf{0.537} & 0.498 \\
\midrule
\multirow{2}{*}{Dataset 3} & LSTM & \textbf{1.256} & -0.546 & 8.982 & 0.249 & -0.546 & -0.045 \\
& Transformers & 1.446 & \textbf{-0.059} & \textbf{7.493} & \textbf{0.208} & \textbf{-0.069} & \textbf{0.320} \\
\midrule
\multirow{2}{*}{Dataset 4} & LSTM & 0.274 & 0.053 & 18.988 & 0.256 & 0.040 & 0.321 \\
& Transformers & \textbf{0.229} & \textbf{0.250} & \textbf{16.931} & \textbf{0.228} & \textbf{0.247} & \textbf{0.602} \\
\midrule
\multirow{2}{*}{Dataset 5} & LSTM & \textbf{0.368} & 0.797 & 6.767 & 0.140 & 0.796 & 0.762 \\
& Transformers & 0.444 & \textbf{0.837} & \textbf{6.412} & \textbf{0.133} & \textbf{0.814} & \textbf{0.797} \\
\midrule
\multirow{2}{*}{Dataset 6} & LSTM & 0.107 & 0.788 & 2.957 & 0.147 & 0.686 & 0.865 \\
& Transformers & \textbf{0.103} & \textbf{0.893} & \textbf{2.352} & \textbf{0.106} & \textbf{0.835} & \textbf{0.957} \\
\bottomrule
\end{tabular}
\end{tiny}
\begin{flushleft}
\textsuperscript{a} E.V.: Explained Variance.  
\textsuperscript{b} N. RMSE: Normalized RMSE.  
\textsuperscript{c} S. C.: Spearman Correlation.  
\end{flushleft}
\end{table}

\begin{table}[h]
\centering
\caption{Performance Results - Cold Start Delay: Dataset (Frequency: Minute)}
\begin{tiny} 
\begin{tabular}{llcccccc}
\toprule
\textbf{Functions} & \textbf{Model} & \textbf{sMAPE} & \textbf{E.V.\textsuperscript{a}} & \textbf{RMSE} & \textbf{N. RMSE\textsuperscript{b}} & \textbf{R2 Score} & \textbf{S. C.\textsuperscript{c}} \\
\midrule
\multirow{2}{*}{Dataset 7} & LSTM & 0.333 & 0 & 4.163 & 0.143 & -0.009 & 0.001 \\
& Transformers & \textbf{0.201} & \textbf{0.717} & \textbf{2.368} & \textbf{0.081} & \textbf{0.700} & \textbf{0.659} \\
\midrule
\multirow{2}{*}{Dataset 8} & LSTM & \textbf{1.187} & -0.225 & 15.601 & 0.288 & -0.330 & -0.062 \\
& Transformers & 1.371 & \textbf{0.274} & \textbf{12.360} & \textbf{0.228} & \textbf{0.141} & \textbf{0.441} \\
\midrule
\multirow{2}{*}{Dataset 9} & LSTM & 0.096 & -0.183 & 53.663 & 0.177 & -0.277 & 0.189 \\
& Transformers & \textbf{0.096} & \textbf{0.193} & \textbf{51.213} & \textbf{0.169} & \textbf{-0.164} & \textbf{0.461} \\
\midrule
\multirow{2}{*}{Dataset 10} & LSTM & 0.166 & 0.747 & 196.030 & 0.119 & 0.738 & 0.753 \\
& Transformers & \textbf{0.128} & \textbf{0.805} & \textbf{167.645} & \textbf{0.101} & \textbf{0.802} & \textbf{0.777} \\
\midrule
\multirow{2}{*}{Dataset 11} & LSTM & 0.425 & 0.458 & 3.518 & 0.152 & 0.450 & 0.704 \\
& Transformers & \textbf{0.412} & \textbf{0.522} & \textbf{3.236} & \textbf{0.140} & \textbf{0.522} & \textbf{0.731} \\
\midrule
\multirow{2}{*}{Dataset 12} & LSTM & 0.108 & 0.813 & 3.060 & 0.122 & 0.711 & 0.898 \\
& Transformers & \textbf{0.043} & \textbf{0.958} & \textbf{1.179} & \textbf{0.047} & \textbf{0.958} & \textbf{0.974} \\
\bottomrule
\end{tabular}
\end{tiny}
\begin{flushleft}
\textsuperscript{a} E.V.: Explained Variance.  
\textsuperscript{b} N. RMSE: Normalized RMSE.  
\textsuperscript{c} S. C.: Spearman Correlation.  
\end{flushleft}
\end{table}

In sum, cold start delay is markedly reduced thanks to the Transformer’s predictive precision, enabling a more responsive and cost-effective environment for HTTP-triggered functions.

\subsection{Reducing Cold Start Frequency}
The frequency of cold starts indicates how often a function restarts from a cold state, directly impacting resource efficiency. We tested two main configurations:
\begin{itemize}
    \item Default (Static) configuration : by default, OpenWhisk keeps a container idle for a fixed 10-minute window. This static approach does not adapt to varying workloads. If demand patterns shift (bursts of invocations followed by long idle periods), the platform either keeps containers around too long (wasting resources) or shuts them down too soon (leading to more cold starts).
    \item Adaptive configuration using Transformers : our model dynamically adjusts the container inactivity window based on predicted demand intervals. The idle time is tailored to each function’s actual usage pattern, resulting in a more flexible, real-time approach.
\end{itemize}

To evaluate frequency-related improvements, we selected 6 different invocation patterns relevant to this problem, each representing distinct usage profiles (sporadic vs. semi-regular vs. bursty traffic). Two key metrics were observed :
\begin{itemize}
    \item Idle-Container Window (Figure 7) : under the default configuration, is always 10 minutes, regardless of actual demand. Under the Transformers approach, the inactivity window adapts in real time. For instance, in one dataset (call it “dataset 13”), the window varied from 14 to 22 minutes, indicating relatively stable usage with predictable idle times. Another dataset (e.g., “dataset 14”) showed a narrower range (7 to 15 minutes), reflecting more frequent usage spikes. In highly irregular datasets, such as “dataset 16,” the window ranged from 5 to 141 minutes, demonstrating the platform’s need for maximum flexibility to avoid unnecessary cold starts.
    \begin{figure*}
        \centering
        \includegraphics[width=1\linewidth]{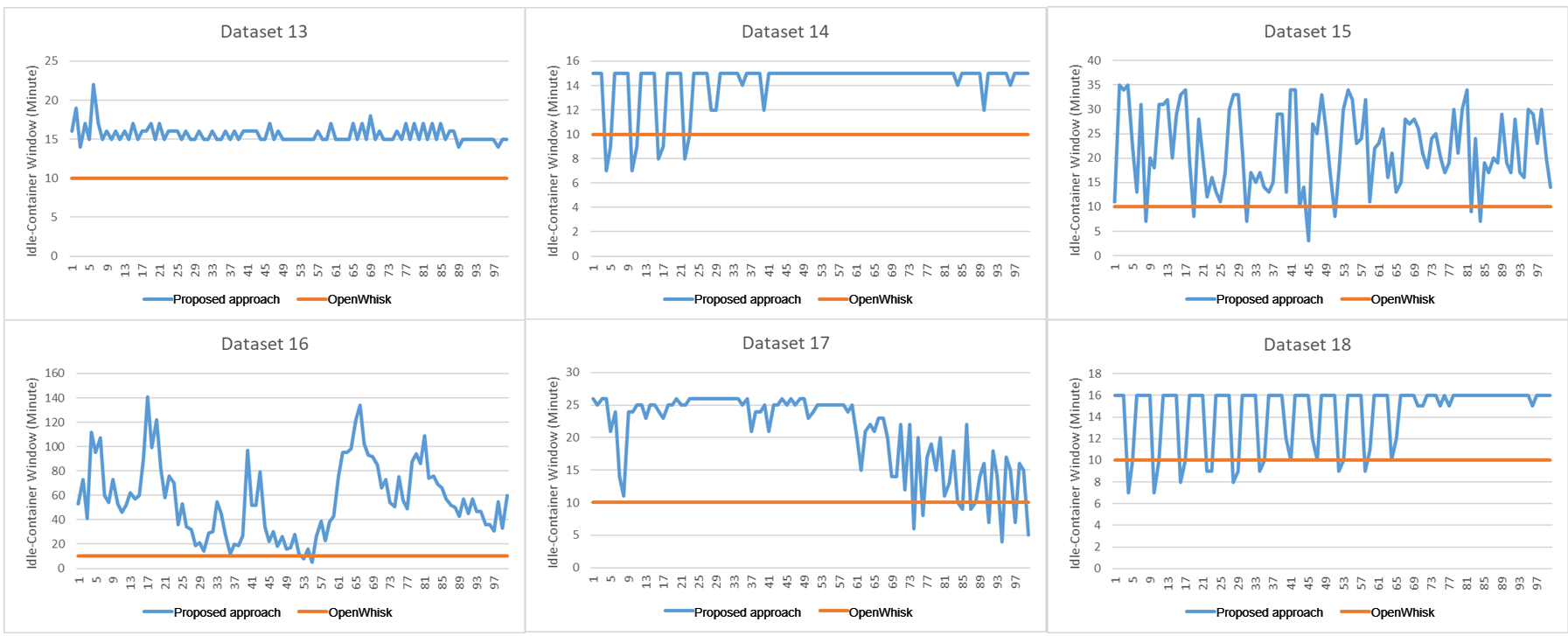}
        \caption{Idle-Container Window}
        \label{fig:enter-label}
    \end{figure*}
    \item Number of Cold Starts (Table III) : with a fixed 10-minute window, the number of cold starts was consistently high across all tested datasets. For example, one dataset recorded 100 cold starts for 100 invocations, implying every request triggered a new container. Others ranged from 66 to 72 cold starts, all of which are suboptimal. By contrast, with dynamic adjustment via the Transformer model, cold starts dropped significantly ; In one dataset, cold starts fell from 100 to 45 (a 55\% reduction). Another dataset saw an even more dramatic reduction from 66 to 14 (79\% reduction). Similar improvements were observed across the board, with reductions of 65\% or more in some cases.
\end{itemize}

This adaptive approach to container management clearly outperforms the static configuration. Specifically : efficient resource utilization, reduced cold start incidence, scalability and flexibility. By using Transformers to predict each function’s unique demand profile, we can dynamically optimize container idle times, substantially lower cold start frequency, and improve user experience.

\begin{table}[!t]
    \centering
    \caption{Comparison of Inactive Container Window Variations and Number of Cold Starts on OpenWhisk with and without Associated Transformers Model}
    \begin{tabular}{|p{1.5cm}|p{2cm}|p{2cm}|p{1.5cm}|}
        \hline
        \textbf{Function} & \textbf{Platform} & \textbf{ICW\textsuperscript{a} Variation} & \textbf{CS\textsuperscript{b} /100} \\
        \hline
        Dataset 13 & OpenWhisk & fixed = 10 min & 100 \\
        & \textbf{OW\textsuperscript{c} + Transf\textsuperscript{d}} & \textit{14-22 min} & \textbf{45} \\
        \hline
        Dataset 14 & OpenWhisk & fixed = 10 min & 66 \\
        & \textbf{OW\textsuperscript{c} + Transf\textsuperscript{d}} & \textit{7-15 min} & \textbf{14} \\
        \hline
        Dataset 15 & OpenWhisk & fixed = 10 min & 72 \\
        & \textbf{OW\textsuperscript{c} + Transf\textsuperscript{d}} & \textit{3-35 min} & \textbf{45} \\
        \hline
        Dataset 16 & OpenWhisk & fixed = 10 min & 71 \\
        & \textbf{OW\textsuperscript{c} + Transf\textsuperscript{d}} & \textit{5-141 min} & \textbf{25} \\
        \hline
        Dataset 17 & OpenWhisk & fixed = 10 min & 62 \\
        & \textbf{OW\textsuperscript{c} + Transf\textsuperscript{d}} & \textit{4-26 min} & \textbf{38} \\
        \hline
        Dataset 18 & OpenWhisk & fixed = 10 min & 66 \\
        & \textbf{OW\textsuperscript{c} + Transf\textsuperscript{d}} & \textit{7-16 min} & \textbf{17} \\
        \hline
    \end{tabular}
    \begin{flushleft}
        \textsuperscript{a} ICW: Idle Container Window. \\
        \textsuperscript{b} CS: Cold start. \\
        \textsuperscript{c} OW: OpenWhisk. \\
        \textsuperscript{d} Transf: Transformers. \\
    \end{flushleft}
    \label{tab:cs_comparison}
\end{table}

By combining DBSCAN clustering to segment functions by invocation pattern, Transformer-based predictions for fine-grained scheduling, and dynamic container management in OpenWhisk, this approach drastically reduces both cold start delay and cold start frequency, ultimately improving the reliability and scalability of serverless applications.

\subsection{General Discussion of Results and Implications for FaaS Environments}
The Transformer model demonstrates significant effectiveness in managing cold starts, outperforming LSTM in two critical aspects: delay and frequency. Key advantages can be summarized as follows :
\begin{itemize}
    \item Reduced Overall Latency : by minimizing cold start delay, the Transformer enables near-instant execution for frequently invoked functions, essential for low-latency services ;
    \item Lower Infrastructure Costs : reduced cold start frequency ensures optimal use of cloud resources, avoiding excessive costs linked to repeated startups ;
    \item Enhanced Responsiveness in Dynamic Environments : thanks to its attention mechanism and efficient temporal relationship processing, the Transformer delivers accurate predictions even in unstable workload environments, crucial for dynamic cloud applications.
\end{itemize}

These results highlight the practical implications of using Transformers in FaaS architectures. By optimizing both prediction accuracy and resource efficiency, our model provides a robust solution for highly responsive and scalable cloud services.

Transformers represent a significant advancement in cold start management for serverless environments. Their prediction and resource optimization capabilities improve both function responsiveness and resource consumption efficiency. Based on these findings, FaaS environments can leverage this architecture to enhance user experience while reducing costs, making it an ideal solution for modern cloud computing.

\section{Model limitations and edge cases}
Although our Transformer-based approach has generally demonstrated superior performance in reducing cold start delays and frequency, certain configurations and datasets reveal less satisfactory results, sometimes even underperforming compared to the LSTM model. This section details these specific cases and associated metrics to identify the model’s limitations.

\subsection{Cold Start delay reduction – invocations per minute}
When analyzing predictions at one-minute intervals, we observed three datasets (Datasets 19, 20, and 21) where the Transformer model occasionally underperformed compared to the LSTM model. A summary of the results is in Table IV.
\begin{table}[h]
\centering
\caption{Performance Results - Cold Start Delay: Minute-Level}
\begin{tabular}{llllllll}
\toprule
\textbf{Func.} & \textbf{Model} & \textbf{sMAPE} & \textbf{E.V.\textsuperscript{a}} & \textbf{RMSE} & \textbf{NRM\textsuperscript{b}} & \textbf{R2 Sc.} & \textbf{S. C.\textsuperscript{c}} \\
\midrule
\multirow{2}{*}{Data19} & LSTM & \textbf{0.837} & \textbf{0.045} & \textbf{1.436} & \textbf{0.239} & \textbf{-0.105} & \textbf{0.448} \\
& Transf. & 1.437 & -0.385 & 1.784 & 0.297 & -0.686 & 0.134 \\
\midrule
\multirow{2}{*}{Data20} & LSTM & \textbf{0.142} & \textbf{0.102} & \textbf{13.651} & \textbf{0.206} & \textbf{0.039} & \textbf{0.385} \\
& Transf. & 0.139 & 0.024 & 13.876 & 0.210 & 0.007 & 0.176 \\
\midrule
\multirow{2}{*}{Data21} & LSTM & \textbf{0.641} & \textbf{0.144} & \textbf{2.384} & \textbf{0.149} & \textbf{0.136} & 0.222 \\
& Transf. & 0.648 & 0.094 & 2.495 & 0.155 & 0.081 & \textbf{0.236} \\
\bottomrule
\end{tabular}
\begin{flushleft}
\textsuperscript{a} E.V.: Explained Variance.  
\textsuperscript{b} N. RMSE: Normalized RMSE.  
\textsuperscript{c} S. C.: Spearman Correlation.  
\end{flushleft}
\end{table}

\subsection{Cold Start delay reduction – invocations per hour}
For invocations aggregated per hour (Datasets 22, 23, and 24), the comparison also reveals cases where the Transformer model does not achieve the expected performance (see Table V).
\begin{table}[h]
\centering
\caption{Performance Results - Cold Start Delay: Hourly}
\begin{tiny}
\begin{tabular}{llcccccc}
\toprule
\textbf{Functions} & \textbf{Model} & \textbf{sMAPE} & \textbf{E.V.\textsuperscript{a}} & \textbf{RMSE} & \textbf{N. RMSE\textsuperscript{b}} & \textbf{R2 Score} & \textbf{S. C.\textsuperscript{c}} \\
\midrule
\multirow{2}{*}{Dataset 22} & LSTM & \textbf{0.029} & \textbf{0.940} & \textbf{0.525} & \textbf{0.087} & \textbf{0.929} & \textbf{0.944} \\
& Transformers & 0.031 & 0.923 & 0.582 & 0.097 & 0.913 & 0.865 \\
\midrule
\multirow{2}{*}{Dataset 23} & LSTM & \textbf{0.507} & \textbf{0.091} & \textbf{1.962} & \textbf{0.245} & \textbf{-0.102} & \textbf{0.383} \\
& Transformers & 0.686 & -0.203 & 2.257 & 0.282 & -0.203 & 0.395 \\
\midrule
\multirow{2}{*}{Dataset 24} & LSTM & 0.829 & \textbf{0.418} & \textbf{1.375} & \textbf{0.196} & \textbf{0.418} & \textbf{0.679} \\
& Transformers & 0.829 & 0.204 & 1.573 & 0.224 & 0.204 & 0.491 \\
\bottomrule
\end{tabular}
\end{tiny}
\begin{flushleft}
\textsuperscript{a} E.V.: Explained Variance.  
\textsuperscript{b} N. RMSE: Normalized RMSE.  
\textsuperscript{c} S. C.: Spearman Correlation.  
\end{flushleft}
\end{table}

\subsection{Cold Start frequency reduction}
Results related to cold start frequency management also highlight some shortcomings of the Transformer-based adaptive model. The comparisons between the default OpenWhisk configuration and the OpenWhisk + Transformers configuration yield the following results (see table VI).
\begin{table}[ht]
    \centering
    \caption{Comparison of Inactive Container Window Variations and Number of Cold Starts on OpenWhisk with and without Transformers Model}
    \begin{tabular}{|p{1.5cm}|p{2cm}|p{2cm}|p{1.5cm}|}
        \hline
        \textbf{Function} & \textbf{Platform} & \textbf{ICW\textsuperscript{a} Variation} & \textbf{CS\textsuperscript{b} /100} \\
        \hline
        Dataset 25 & OpenWhisk & fixed = 10 min & \textbf{33} \\
        & \textbf{OW\textsuperscript{c} + Transf\textsuperscript{d}} & 1-25 min & 40 \\
        \hline
        Dataset 26 & OpenWhisk & fixed = 10 min & \textbf{71} \\
        & \textbf{OW\textsuperscript{c} + Transf\textsuperscript{d}} & 2-23 min & 76 \\
        \hline
        Dataset 27 & OpenWhisk & fixed = 10 min & \textbf{42} \\
        & \textbf{OW\textsuperscript{c} + Transf\textsuperscript{d}} & 6-12 min & 47 \\
        \hline
    \end{tabular}
    \begin{flushleft}
        \textsuperscript{a} ICW: Idle Container Window. \\
        \textsuperscript{b} CS: Cold Start. \\
        \textsuperscript{c} OW: OpenWhisk. \\
        \textsuperscript{d} Transf: Transformers. \\
    \end{flushleft}
    \label{tab:cs_freq}
\end{table}

These observations show that in some cases, the dynamic container inactivity window adjustment algorithm failed to optimize resource management, leading to an increase in cold starts. 

In summary, while Transformers provide notable improvements in many scenarios, these negative results highlight situations where the model has not met expectations. This calls for targeted improvements to ensure consistently strong performance across all use cases.

\section{Conclusion}
In this research, we proposed an innovative approach to mitigate cold start issues in Function-as-a-Service (FaaS) environments, relying on advanced deep learning models, specifically Transformers. Our contributions are summarized as follows: we developed a method that anticipates and reduces cold start delays, significantly improving the responsiveness of cloud functions; our Transformer-based model outperformed traditional approaches, particularly in environments with high workload variability, by enabling more accurate predictions of future invocations and optimized container management; in terms of cold start delay reduction, our models demonstrated enhanced performance, although certain limitations, such as the granularity of invocation data, restricted the prediction accuracy in cases with short intervals.

The results of this study have significant implications for both research and practical applications in the field of cloud computing. By adopting Transformer models for the prediction and proactive management of cold starts, our work paves the way for more robust and responsive solutions in serverless environments. These improvements are particularly beneficial for applications where low latency is critical, such as real-time systems and highly interactive services. However, the observed limitations—particularly the temporal aggregation of data and the absence of certain data due to privacy concerns—highlight the importance of adapting these models to the specificities of platforms and workloads. This work also provides a solid foundation for future optimizations, particularly in reducing operational costs related to cloud resources.

The encouraging results obtained in this study suggest several areas for further exploration :
\begin{itemize}
    \item Exploration of new Transformer architectures : testing and adapting more recent and specialized Transformer architectures could provide additional improvements for cold start mitigation, particularly in terms of speed and accuracy ;
    \item Integration of reinforcement learning : adding reinforcement learning techniques would allow for dynamic, real-time cloud resource management, ensuring more responsive and optimal resource allocation ;
    \item Advanced resource management optimization : more sophisticated memory management mechanisms could be incorporated to further balance performance and operational costs. For example, the implementation of adaptive algorithms for maintaining containers in memory ;
    \item Inclusion of cross-platform data : by incorporating usage data from various FaaS platforms, as well as application metrics, it would be possible to improve the accuracy of models and tailor them more specifically to the nuances of different usage contexts.
\end{itemize}
In conclusion, our research highlights the potential of deep learning models to address the specific challenges of FaaS architectures while laying the groundwork for future innovations in proactive serverless resource management \cite{agarwal_reinforcement_2021, agarwal_reinforcement_2023}.

\bibliography{main}
	\bibliographystyle{ieeetr}
\vfill

\end{document}